\documentclass{statsoc}
\usepackage[cm]{fullpage}
\usepackage[figuresright]{rotating}
\usepackage{epstopdf}
\usepackage{url}
\usepackage{color}
\usepackage{tikz}
\usetikzlibrary{shapes,arrows}
\usetikzlibrary{calc,positioning}
\usepackage{multirow}
\usepackage[caption=false]{subfig}
\usepackage{graphicx}
\usepackage{enumerate}
\usepackage[titletoc,title]{appendix}
\usepackage{threeparttable, tablefootnote}

\usepackage{amsmath, amsthm, amssymb}
\usepackage{array,colortbl,xcolor,tabularx}
\usepackage{natbib}
\usepackage{enumitem}
\usepackage{xr}
\usepackage{threeparttable}

\newcommand\independent{\protect\mathpalette{\protect\independenT}{\perp}}
\def\independenT#1#2{\mathrel{\rlap{$#1#2$}\mkern2mu{#1#2}}}

\newcommand{\tX}{\textbf{X}}
\newcommand{\tx}{\textbf{x}}
\newcommand{\sgn}{\text{sgn}}
\newcommand{\ys}{Y^{\ast}}

\title[Biomarker selection for treatment recommendation]{Selecting Biomarkers for building optimal treatment \\selection rules using Kernel Machines}
\author[Sayan Dasgupta {\it et al.}]{Sayan Dasgupta and Ying Huang}
\address{Vaccine and Infectious Disease Division, Fred Hutchinson Cancer Research Center,
Seattle,
USA.}
\email{sdasgup2@fredhutch.org}

\begin{document}

\setlength{\abovedisplayskip}{2pt}
\setlength{\belowdisplayskip}{2pt}




\begin{abstract}
Optimal biomarker combinations for treatment-selection can be derived by minimizing total burden to the population caused by the targeted disease and its treatment. However, when multiple biomarkers are present, including all in the model can be expensive and hurt model performance. To remedy this, we consider feature selection in optimization by minimizing an extended total burden that additionally incorporates biomarker measurement costs. Formulating it as a 0-norm penalized weighted-classification, we develop various procedures for estimating linear and nonlinear combinations. Through simulations and a real data example, we demonstrate the importance of incorporating feature-selection and marker cost when deriving treatment-selection rules.
\end{abstract}


\maketitle

\section{Introduction.}
A considerable amount of recent biometric research is being conducted in the `personalized medicine' framework, because it has been well accepted now that heterogeneity can exist among individual subjects' response to treatment in many disease settings. 
The characteristics contributing to this heterogeneity may include patient demographics, genetic/genomic information or other biological markers, henceforth referred to as treatment-selection biomarkers. These biomarkers can be effectively used to select optimal therapies for individuals in order to optimize this clinical outcome. Also it is important to remember that a single biomarker may not sufficiently explain this heterogeneity, and multiple biomarkers may need to be combined to build the correct statistical framework to optimize the process of treatment selection. 

The direct approach to identifying these optimal marker combinations in treatment selection involves parametric modeling of the disease risk conditional on biomarkers, treatment assignment and other baseline patient characteristics, and recommending treatment assignments based on whether the predicted risk under treatment is lower than the predicted risk under no treatment. This framework was first introduced by \citet{Song2004}, and studied extensively in \citet{Foster2011, Qian2011, Lu2013}.
Treatment selection rules based on parametric risk models rely heavily on the correct specification of this model, which is often challenging given the complexity of biological mechanisms. An alternate, much more robust approach is to build optimization algorithms to minimize (or maximize) a desired criterion. This criterion, often called the objective function, is formulated based on relevant goals for treatment selection, and the optimization algorithm tries to find the best marker combination that optimizes it. These methods (also called indirect approaches) can be made completely nonparametric and assumption free and much more robust to model misspecifications. 
\citet{Zhang2012a, Zhang2012b} proposed finding the optimal marker combination within a pre-specified class by optimizing an estimator of the overall population mean outcome. \citet{Zhao2012} approached the same as an outcome-weighted learning problem and derived optimal treatment-selection rules using a weighted support vector machine method.

In \citet{Huang2014}, the authors proposed a new method to identify linear and nonlinear marker combinations by directly optimizing a targeted criterion function in the manner of \citet{Zhang2012a} and \citet{Zhao2012}. However, the targeted criterion differed from others as adverse side-effects and/or cost incurred by the treatment were considered along with the event rates of the targeted disease in establishing the objective function. This was a crucial extension, as reducing safety events or cost of administering the treatment is often a valid policy goal \citep[see][]{Vickers2007}. The authors formulated the optimization problem as minimization of a weighted sum of 0-1 loss, and used the ramp loss as an approximation of the 0-1 loss.
In this article, we extend the idea of \citet{Huang2014} in creating an augmented target criterion that 
takes on the additional challenge of controlling for the number of markers in the risk model as well. This is crucial for two vital reasons:
(a) Measuring biomarkers can be expensive with respect to both time and money, and possibly invasive too. 
It is therefore of significant interest to limit the number of biomarkers that need to be collected for an individual to selecting his/her optimal treatment. 
(b) As discussed earlier, our original problem requires us to find optimal biomarker combinations to explain the disease response heterogeneity in individuals, but finding this optimal combination can be difficult in the presence of redundant markers that do not contribute to treatment selection. This may lead to overfitting and result in overly complex selection rules that have poor prediction performance. 
One way to deal with this `curse of dimensionality' is through marker selection. Recently a lot of attention has been directed to effect modifier selection in precision medicine. For example, in \citet{Zhao2017}, the authors studied selective inference in effect modification models via LASSO; in \citet{Liang2017}, the authors constructed sparse decision rules in the context of concordance-assisted learning; and in \citet{Shi2018}, the authors proposed a penalized multi-stage A-learning algorithm for deriving the optimal dynamic treatment regime. Although these feature selection methods are all relevant to the broad genre of precision medicine, each of them target a very specific problem, which are quite different from our objective of penalized optimization of the single stage targeted objective function, where still only limited research exists: some insight into marker selection in treatment recommendation problems were briefly studied by \citet{Huang2015} and \citet{Zhou2015}, where both targeted minimization of disease rate in treatment selection.

To address these issues along with the original goals of treatment selection, we reformulate the criterion for treatment selection by associating a cost for measuring each biomarker in creating the optimal rule. The optimization problem can be written as minimization of a weighted sum of 0-1 loss, but with a $L_0$ penalty added for the number of markers in the model. In this article, we adopt the usual hinge loss convex relaxation of the 0-1 loss and perform a comprehensive investigation and comparison of various algorithms for solving this optimization problem.
In linear support vector machines, $L_0$ feature extraction is a well-researched problem \citep[see][]{Bradley1998, Weston2003, Mangasarian2006, Huang2010}, although it is a much more challenging problem in nonlinear support vector machines \citep[see][]{Mangasarian2007}. However, it is worth noting that our setting differs from a simple classification format in two vital aspects:  (a) although the treatment selection objective can be rewritten into a (weighted) classification problem (as shown in Section \ref{sec:methods}), it is still in essence a fundamentally different problem from classification, and feature selection techniques in SVMs have not been studied under this context, and (b) weighted SVM is a more complicated optimization problem than the standard SVM, where the constraint on each support vector varies according to the weight associated with it, and research into feature extraction under this setting has also been fairly limited till now. Some work has been done to extend the SCAD penalty with the weighted linear support vector machines with special forms of such weights \citep[see][]{Jung2013}, but beyond that, there hasn't been any targeted investigation of such as per our knowledge. These two reasons make these explorations vitally important. 

Thus, the biggest contribution of this article is to combine the methodological advances in two different areas of research, advances in indirect approaches to treatment selection through kernelized methods and those in feature selection techniques in nonparametric statistical learning methods like SVMs. We believe that combining these two areas of research is a novel methodological formulation in itself. Additionally each of these feature selection methods has been translated from the standard SVM framework to the weighted SVM formulation to match the proposed objective. Moreover, adding a penalty for the number of markers in the model redefines the objective function as well, which has been used in conjunction with these feature selection techniques - for example, in choosing tuning parameters for a given feature selection method, we propose the use of a generalized cross validation (GCV) technique that utilizes the whole objective criterion including the penalty for the number of markers. 

The article continues in the following manner: In Section \ref{sec:methods}, we establish the problem of minimizing the total disease, treatment, and marker measurement cost in treatment-selection, and discuss marker selection in conjunction with treatment selection. We briefly discuss various methods built for support vector machines, adapted specifically for the weighted classification setup, for deriving the best linear and nonlinear marker combinations in order to optimize the desired objective function. In Section \ref{sec:simu}, we set up different simulation examples to test the strength of these linear and nonlinear feature selection methods, modified for our setting, and discuss the results. Then in Section \ref{sec:real_data}, we illustrate the application of our methods using a real data example from an HIV vaccine trial. And finally, in Section \ref{sec:disc}, we discuss our findings regarding the use of these feature selection methods and the impact of incorporating marker measurement cost into treatment selection, and link to
additional materials are presented in the Supplementary Section \ref{sec:supp}.

\section{Methods.} \label{sec:methods}
In this article, we consider the problem of finding optimal treatment selection rules for a binary clinical outcome $Y$ (0 for nondiseased and 1 for diseased) based on a set of $p \geq 1$ candidate markers collected from an individual's biological characteristics (the solution we propose is applicable to clinical outcome measured in continous scale as well). Denote this set of candidate markers by $\textbf{X}_p=\{X_1,\dots,X_p\}$. For a given subset of markers $\textbf{X}\subseteq\textbf{X}_p$, we consider marker based treatment selection rules of the form $A(\textbf{X})=I(f(\textbf{X})>0)$, where $f\in\mathcal{F}_{\textbf{X}}$ and $\mathcal{F}_{\textbf{X}}$ denotes a class of functions spanning over $\textbf{X}$, and $I(\cdot)$ is the indicator function. The above then translates to the rule: $A = 0$ for not treating, and $A = 1$ for treating. The treatment-selection benefit of a decision rule  $A(\textbf{X})$ can be quantified by $E_{A(\textbf{X})}(Y)$, the expected disease rate in the population as a result of treatment selection based on $A(\textbf{X})$ \citep{Song2004, Qian2011}. This measure has been widely accepted as a crucial metric in recent literature \citep{Zhao2012, Zhang2012a}. 
Although $E_{A(\textbf{X})}(Y )$ characterizes the burden of disease upon the population, one may be concerned with additional burden associated with a treatment regimen, 
such as its side effects and/or the monetary cost of its implementation. Hence it might be preferable to search for treatment-selection strategies that take these aspects into consideration as well. 
\citet{Huang2014} proposed to incorporate additional burden associated with a treatment regimen, such as due to its side effects of monetary cost, by pre-specifying a treatment/disease harm ratio such that each burden type can be put on the same scale. Following the decision-theoretic framework of \citet{Vickers2007}, let $\delta_1$ be a pre-specified ratio of the burden per treatment relative to the burden per disease event, and let $Y (1)$ and $Y (0)$ indicate the potential disease outcome if a subject were to receive or not receive the treatment. Then the total burden due to disease and treatment for $A(\textbf{X})$, represented in the unit of burden per disease event \citep[see][]{Huang2014} is given as $E_{A(\textbf{X})}(Y) + E\{\delta_1 \times A(\textbf{X})\}$
$=\sum_{a=0}^1 E\left[Y(a)\times I\{A(\textbf{X}=a)\}\right]+\delta_1\times E\{A(\textbf{X})\}$.
And the best treatment-selection rule is derived as the one that minimizes this total burden.

In this article we further take into consideration the cost of measuring biomarkers in deriving the optimal treatment-selection rule, under the expectation that inclusion of the burden of measuring biomarkers in the criterion function can lead to the derivation of more cost-effective treatment-selection rules in the sense of achieving desired public health impact under the guidance of a parsimonious biomarker panel. We make the following assumptions: (i) measurement of each biomarker induces roughly equal cost, and (ii) the cost of measuring one biomarker is $\delta_2$ times the burden per disease event. Then the total burden due to disease, treatment, and biomarker measurement for a treatment-selection rule $A(\textbf{X})$ can be represented in the unit of the burden per disease event as
\begin{align}\label{eq:theta2}
\theta &= E_{A(\textbf{X})}(Y) + E\{\delta_1 \times A(\textbf{X})\}+\delta_2\times \dim(\textbf{X})\nonumber\\
&=\sum_{a=0}^1 E\left[Y(a)\times I\{A(\textbf{X})=a\}\right]+\delta_1\times E\{A(\textbf{X})\}+\delta_2\times \dim(\textbf{X}).
\end{align}
 \textbf{Note:} Although we assume a roughly equal cost for each biomarker for simplicity in this paper, it is straightforward to extend it to a setting where each biomarker has a different cost.  Biomarkers which are absolutely essential to collect can be considered to have no cost at all, or there may be several groups of biomarkers, such that each group has a different cost depending on its burden and importance for the population of interest. In such situations, the penalty $\delta_2 \times \dim(\textbf{X})$ can be replaced by $\sum_{i=1}^{\dim(\textbf{X})} \delta_{2,i}$ where $\delta_{2,i} \geq 0$ is the cost associated with the biomarker $X_i$, and our methods can still be applied as described with only minor modifications.

We propose to derive an optimal treatment-selection rule by minimizing this quantity $\theta$. Suppose there exists an `optimal' or `correct' set of $p_0$ biomarkers $\textbf{X}_0$ within the list of biomarkers measured, such that $\textbf{X}_0=\{X^{\ast}_1,\dots,X^{\ast}_{p_0}\}\in \textbf{X}_p$, which when combined through an oracle rule $f_0$ minimizes $\theta$. Thus, we can write the above problem as,
\begin{align}\label{eq:theta0}
f_0=\displaystyle{\arg \min_{\textbf{X} \subseteq \textbf{X}^{p}}\min_{f\in\mathcal{F}_{\textbf{X}}} } \theta
\end{align}
Note that under this setup, $f_0\in \mathcal{F}_{\textbf{X}_0}$. In practice, it is important to have a sensible way to specify the values or the range of values for $\delta_1$ and $\delta_2$. 
Choice of these cost ratios can be facilitated based on information of the monetary cost for controlling the targeted disease, for applying the treatment, and for biomarker measurement, as in our data example of making recommendation for HIV vaccine, presented later in Section \ref{sec:real_data}.   

Now imagine data from a two-arm randomized trial with the treatment indicator $T$ taking values 0 and 1, to refer being untreated and treated, respectively. Let $n_0$ and $n_1$ indicate the number of subjects in the untreated and treated arms, respectively. Thus we have i.i.d. samples of the form $\{Y_i, \textbf{X}_i, T_i\}$ for $i = 1,\dots, n$ with $n = n_0 + n_1$. As in \citet{Huang2014}, we assume: (i) stable unit treatment value (SUTVA) \citep{Rubin1980} and consistency: $Y(0)$, $Y(1)$ of one subject is independent of the treatment assignments of other subjects, and given the treatment a subject actually received, a subject's potential outcomes equal the observed outcomes; (ii) ignorable treatment assignments assumption: $T \independent Y (0), Y (1)|\textbf{X}$. Assumption (i) is plausible in trials where participants do not interact with one another and assumption (ii) is ensured by randomization. Under the above assumptions, it can be shown that,
{\small\begin{align}\label{eq:theta3}
\theta&=E\left[Y\times\{1-A(\tX)\}|T=0\right]+E\left[Y\times A(\tX)|T=1\right] + \delta_1\times P\{A(\tX)=1\}+\delta_2\times \dim(\tX)\nonumber\\
&=E(Y|T=0)-E\left[A(\tX)\times \{Risk_0(\tX)- Risk_1(\tX)-\delta_1\}\right] +\delta_2\times \dim(\tX)
\end{align}}
where $Risk_0(\tX) = P(Y = 1|\tX, T = 0)$ and $Risk_1(\tX) = P(Y = 1|\tX, T = 1)$ are the risk of $Y$ conditional on $\tX$ among the untreated and treated, respectively.

\citet{Huang2014} showed that when $\delta_2=0$, an optimal rule $A(\tX)$ can be obtained as $A(\tX) = 1$ if $Risk_0(\tX) -
Risk_1(\tX) > \delta_1$, and $A(\tX) = 0$ otherwise. But when $\delta_2$ is positive, such a strategy alone wouldn't exactly work as we need to also control the number of markers in the model. Since the quantity $\delta_2$ imposes an $L_0$ penalty on the set of markers $\tX$, one ad hoc method may be to use standard regression-based methods with $L_0$/$L_1$ penalization to estimate the quantities $Risk_0(\tX)$ and $Risk_1(\tX)$ and then derive the optimal treatment-selection rule as $A(\tX) = I\{Risk_1(\tX) - Risk_0(\tX) > \delta_1\}$. But note that the above procedure may lead to suboptimal rules with respect to our goal of minimizing $\theta$. 

Alternatively, following the strategy in \citet{Zhang2012a}, \citet{Zhao2012} and \citet{Huang2014}, we can consider a class of rules for treatment recommendation based on functionals of the form $f(\tX)$ (belonging to some functional class $\mathcal{F}_{\tX}$ on $\tX$), and a given threshold (usually 0). In particular, we let $A(\tX) = I\{f(\tX) > 0\}$ with $I(\cdot)$ the indicator function, $f(\tX) = b +g(\tX)$ with $g(\tX)$ a function of markers $\tX$. Then, assuming randomization does not depend on $\tX$ for simplicity, $\theta$ can be expressed as
$\frac{E\left[Y\times T\times I\{f(\tX)>0\}\right]}{P(T=1)}+\frac{E\left[Y\times (1-T) \times I\{f(\tX)\leq 0\}\right]}{P(T=0)}+ \delta_1\times P\{f(\tX)>0\}+\delta_2\times \dim(\tX)$.
The optimal $f(\tX)$ can be found by minimizing the empirical estimate of $\theta$, that is, 
{\small\begin{align*}
\hat{f}=\arg \min_{\textbf{X} \subseteq \textbf{X}^{p}}\min_{f\in\mathcal{F}_{\textbf{X}}} \left\{\sum_{i=1}^n \{Y_i \times \frac{T_i}{n_1} - Y_i \times \frac{(1-T_i)}{n_0} +\frac{\delta_1}{n} \} \times I\{ f(\tX_i)\leq 0\}+\delta_2 \times \dim(\tX)\right\}.
\end{align*}}


Therefore, we can formulate this problem as the minimization of the sum of a weighted sum of 0-1 loss and a term proportional to the number of biomarkers. That is, $\hat{f}$ can be found as the minimizer of
\begin{align}\label{eq:wt_sum}
\sum_{i=1}^n W_i I\left\{f(\tX_i)\leq 0\right\}/n+\delta_2 \times \dim(\tX),
\end{align}
with the case-specific weight $W_i= W_{1i}=- \left\{	\frac{Y_i \times T_i}{n_1/n}-\frac{Y_i \times (1-T_i)}{n_0/n}+\delta_1\right\}$. 
Other types of weights such as control-specific or case-control mixture weights or their robust substitutes can also be adopted and are discussed in \citet{Huang2014}. For example, the robust substitute for the case-only weights $W^{rob}_{1i}=-\left\{	\frac{Y_i \times T_i}{n_1/n}-\frac{Y_i \times (1-T_i)}{n_0/n}+\frac{\pi-T_i}{\pi}\widehat{Risk}_0+\frac{\pi-T_i}{1-\pi}\widehat{Risk}_1+\delta_1\right\}$,
where $\widehat{Risk_{0}}(\tX)$ and $\widehat{Risk_{1}}(\tX)$ are risk estimates obtained from a working model and $\pi=P(T=1)$. It is worthwhile to note that since the minimization of \eqref{eq:theta3} is equivalent to the minimization of $E[I\{f (\tX) \leq 0\} \times \{Risk_0(\tX) - Risk_1(\tX) - \delta_1\}]$, any consistent estimate of $Risk_0(\tX) - Risk_1(\tX) - \delta_1$ can also serve as weights in \eqref{eq:wt_sum}. 

Note that if we are only interested in combining a set of candidate biomarkers without performing any variable selection, then it is not necessary to include the biomarker measurement cost in the criterion function as in \citet{Huang2014}. However, in the presence of a large number of biomarkers, incorporation of marker cost will impact the features selected in the estimated rule.

\subsection{Treatment selection as a weighted Support Vector Machines problem.}
In this section, we consider the minimization of the regularized loss function \eqref{eq:wt_sum}, conditional on a pre-specified set of weights $W$. Since $\sum_{i=1}^n W_i I\left\{f(\tX_i)\leq 0\right\}/n \propto \sum_{i=1}^n |W_i| I\left\{\sgn(f(\tX_i))\neq \sgn(W_i)\right\}/n$, \eqref{eq:wt_sum} can be reformulated as 
\begin{align}\label{eq:wt_sum2}
\sum_{i=1}^n |W_i| I\left\{\sgn(f(\tX_i))\neq \sgn(W_i)\right\}/n+\delta_2 \times \dim(\tX)
\end{align}
Note that \eqref{eq:wt_sum2} is a weighted classification problem where $Y^{\ast}_i=\sgn(W_i) = \{-1, 1\}$ is the true binary classes, $\sgn\{f (\tX_i)\}$ is the predicted binary class based on $\tX$, and $|W_i|$ is the subject specific weight. This type of problem can be resolved using the weighted support vector machine \citep{Lin2002}, based on $\tX_i$, $Y_i$, and $|W_i|$. Since minimization of a weighted sum of 0-1 loss is non-convex and intractable, we look to replace the 0-1 loss with a convex surrogates; the hinge loss is often used in this context. The hinge loss, given as $h(u) = \max (0, 1 - u)$ has been proven to be a useful surrogate to the classification loss, such that the term $ |W_i| I\left\{\sgn(f(\tX_i))\neq \sgn(W_i)\right\}$ in \eqref{eq:wt_sum2} is replaced by $ |W_i| h\left(f(\tX_i)\times\sgn(W_i)\right)$. 
It is worth noting that the hinge loss penalizes departure of a decision rule from its observed class label, based on the extent of this departure, which the classification loss fails to do. The hinge-loss SVM formulation of the above problem is given as, 
 \begin{align}\label{eq:l0_svm}
\displaystyle{ \min_{\textbf{X} \subseteq \textbf{X}^{p}}\min_{f\in\mathcal{F}_{\textbf{X}}} } \sum_{i=1}^n |W_i| \max(1 - \sgn(W_i) \times f(\tX_i), 0)/n +  \delta_2 \times \dim(\tX)
\end{align}
This is the $L_0$ penalized weighted support vector machines framework. In SVMs, the functional space $\mathcal{F}_{\tX}$ is generally restricted to be a reproducing kernel Hilbert space $\mathcal{H}_{\tX}(k_{\tx})$, represented uniquely by its kernel $k_{\tx}$. 
Before considering the solution to \eqref{eq:l0_svm}, we first review the weighted support vector machines formulation from \citet{Lin2002}, where the square of the Hilbert space norm $\|\cdot\|_{\mathcal{H}_{\tX}}$ is used instead of the $L_0$ norm. For a given subspace $\tX \in \tX^p$, we define it as:
\begin{align}\label{eq:l2_svm}
\displaystyle{\min_{f\in\mathcal{H}_{\tX}} } \sum_{i=1}^n |W_i| \max(1 - \sgn(W_i) \times f(\tX_i), 0)/n +  \lambda \|f\|^2_{\mathcal{H}_{\tX}}
\end{align}
In linear support vector machines, $\mathcal{F}^{\text{lin}}_{\tX}=\{f_{\beta,b_{0}}:f_{\beta,b_{0}}(\textbf{x})=\left\langle \beta, \textbf{x}\right\rangle + b_{0},\: \textbf{x} \in \tX,\: \beta \in \mathbb{R}^{\dim(\tX)},\: b \in \mathbb{R}\}$ is used for optimization, an RKHS with the Euclidean inner product as its kernel function, $k_{\tx}(\textbf{x}_1,\textbf{x}_2)=\left\langle \textbf{x}_1, \textbf{x}_2\right\rangle$. The Hilbert norm $\|f\|^2_{\mathcal{H}_{\tX}}$ in this case becomes the usual Euclidean $L_2$ norm on the linear combination weights $\|\beta\|^2$, and the estimated decision functions $\hat{f}$ can be expressed as a linear combination of the input marker set $\tX$. However, as many optimal marker combination for treatment selection may not be among linear combinations of the input markers, it is important to extend $\mathcal{F}_{\tX}$ to include more complex nonlinear functions. This can be achieved by considering transformations of the feature space through feature maps of the form $\phi(\tx)$, the appropriate RKHS for which is the one with the kernel $k_{\tx}$ satisfying $k_{\tx}(\textbf{x}_1,\textbf{x}_2)=\left\langle \phi(\tx_1), \phi(\tx_2)\right\rangle$. 
Examples of popular nonlinear kernels include the polynomial kernel of $d^{th}$ degree $k_{\tx}(\tx_i, \tx_j) = (1+ \left\langle \tx_i,\tx_j \right\rangle)^d$ and the radial basis function (RBF) kernel $k_{\tx}(\tx_i, \tx_j) = \exp\left(-\gamma\|\tx_i -\tx_j\|^2\right)$ with $\gamma$ as a tuning parameter. 
The resulting SVM solution at a given covariate vector $\tx_0$ is given as $f(\tx_0)=\sum_{j=1}^n\alpha_j k(\tx_0,\tx_j)+\beta_0$, where $\alpha_i \in \mathbb{R}$'s are the trained weights on the support vectors $k(\cdot,\tx_j)$ and $b_0 \in \mathbb{R}$ is the estimated global constant.

\subsection{Weighted versus unweighted support vector machines.} The weighted support vector machines is a well-known technique for classification. First proposed by \citet{Lin2002}, who called it the ``fuzzy support vector machines'' or FSVM, it has been further applied and studied in subsequent works \citep[][see for example]{Fan2005, Yang2007, Zhao2012}. In essence, the weighted support vector machines becomes a very important tool in classification, when some training points are more important than others for the given problem. The main difference in solving the weighted SVM and the standard unweighted SVM lie in the constraints that we put on the support vectors in the dual formulation of the problem. To see this, let us write down the dual of the unweighted SVM below and note the changes that can transform it into a weighted SVM problem. Let us denote $\ys_i=\sgn(W_i)$ and $q_i=|W_i|$, and then see that the unweighted squared Hilbert loss penalized SVM is given as $\min_{f\in\mathcal{H}_{\tX}}  \sum_{i=1}^n \max(1 - \ys_i f(\tX_i), 0)/n +  \lambda \|f\|^2_{\mathcal{H}_{\tX}}$, 
the dual of which is given as, 
\begin{align}\label{eq:dual_unwt}
\max_{\boldsymbol{\alpha}} \sum_{i=1}^n \alpha_i -\frac{1}{2}\sum_{i=1}^n \sum_{j=1}^n \alpha_i\alpha_j\ys_i\ys_j k_{\tX}(\tx_i,\tx_j)\quad \text{subject to }\sum_{i=1}^n \ys_i \alpha_i =0,\: 0\leq \alpha_i\leq C,
\end{align}
for $C=1/(2\lambda n)$. The dual of \eqref{eq:l2_svm} differs from \eqref{eq:dual_unwt} only through the constraints that we put on the $\alpha_i$s. In the unweighted SVM, the upper-bound for each of $\alpha_i$ is the fixed constant $C$, while in the weighted SVM, the upper bound for each $\alpha_i$ is further multiplied by the quantity $q_i$ and thus becomes $C\times q_i$ (that is, $0\leq \alpha_i\leq Cq_i$). Intuitively it just means that $\alpha_i$, the coefficient associated with a given sample, is each constrained differently according to their weight. 

\subsection{Identifying optimal biomarker combinations in the linear space.} As noted before, variable selection is a key motivation in our pursuit to solve \eqref{eq:l0_svm}, which isn't an inherent feature of the support vector machines framework of \eqref{eq:l2_svm}. The Hilbert norm provides some control on the overcomplexification of the estimated functions, but without performing any variable selection. Many authors have proposed to use the $L_1$ and $L_0$ penalty either as a replacement or in conjunction with the $L_2$ norm in the linear SVMs. 
First of all, note that solving \eqref{eq:l0_svm} is combinatorially a very hard problem \citep[see][]{Amaldi1998}, 
but the zero norm is directly related to finding minimal subsets and can provide optimal feature extraction if properly implemented. There are however many feature extraction algorithms for the unweighted linear support vector machines that do not rely on the $L_0$ norm for selection \citep[see for example][]{Mangasarian2006, Zhang2006}.  
To this effect, in this article, we also look at alternate ways to perform feature selection in support vector machines, instead of focusing solely on solving \eqref{eq:l0_svm}. Here, we build our optimization approaches on some of the most commonly used algorithms for unweighted SVMs. That is, we modify each of them to cater to the weighted version of the algorithm, and we will see how it paves the way for directly or indirectly solving \eqref{eq:l0_svm} in the linear space. We now briefly introduce each method below, along with their modified objective function under the weighted SVM setup, while a more detailed description of each is reserved for the Supplementary Section (see Web Appendix \ref{sec:appA}).
\begin{enumerate}
[leftmargin=0.2in,noitemsep]
\item \emph{$L_1$ weighted support vector machines ($L_1$ WSVM)}, that utilizes the $L_1$ norm for penalization instead of the squared Hilbert norm, built upon \citet{Mangasarian2006}, with objective function,
\begin{align}
\min_{\boldsymbol{\beta},\beta_0} \sum_{i=1}^n q_i\max\{1 - \ys_i (\left \langle X_i, \boldsymbol{\beta} \right \rangle+\beta_0), 0\}/n +  \lambda \sum_{j=1}^p |\beta_j|.
\end{align} 
\item  \emph{SCAD} applies the smoothly clipped absolute deviation penalty \citep{Zhang2006} to the WSVM objective function , 
\begin{align}
\min_{\boldsymbol{\beta},\beta_0} \sum_{i=1}^n q_i \max\{1 - \ys_i (\left\langle X_i, \boldsymbol{\beta} \right\rangle +\beta_0), 0\}/n +  \sum_{j=1}^p p_{\lambda}(|\beta_j|),
\end{align}
$p_{\lambda}(|\boldsymbol{\beta}|)= \lambda |\boldsymbol{\beta}|1(|\boldsymbol{\beta}| \leq \lambda) - \frac{|\boldsymbol{\beta}|^2 -2a\lambda|\boldsymbol{\beta}| +\lambda^2}{2(a-1)}1(\lambda < |\boldsymbol{\beta}| \leq a\lambda) +\frac{(a+1)\lambda^2}{2}1(|\boldsymbol{\beta}| > a \lambda)$.
\item \emph{elastic SCAD}, that uses a mixture of the SCAD penalty and the $L_2$ norm,
\begin{align}
\min_{\boldsymbol{\beta},\beta_0} \sum_{i=1}^n q_i \max\{1 - \ys_i (\left\langle X_i, \boldsymbol{\beta} \right\rangle +\beta_0), 0\}/n +  \sum_{j=1}^p p_{\lambda_1}(|\beta_j|) +\lambda_2\|\boldsymbol{\beta}\|^2.
\end{align}
\item \emph{feature selection concave (FSV)}, built on a concave minimization algorithm originally proposed by \citet{Bradley1998}, achieves approximate $L_0$ penalization in weighted SVMs through solving,
\begin{align}
\min_{\boldsymbol{\beta},\beta_0,\boldsymbol{v}} &\frac{\sum_{\ys_i=1} q_i\max\{1 -   \left\langle X_i, \boldsymbol{\beta}\right\rangle - \beta_0, 0\}}{\sum_{i=1}^{n}I(\ys_i= 1)}+\frac{\sum_{\ys_i= -1} q_i\max\{1 -   \left\langle X_i, \boldsymbol{\beta}\right\rangle - \beta_0, 0\}}{ \sum_{i=1}^{n}I(\ys_i= -1)} \nonumber\\ &+ \sum_j \alpha e^{-\alpha v_j^{\ast}(v_j-v_j^{\ast})}\quad
\text{subject to } -\boldsymbol{v} \leq \boldsymbol{\beta} \leq \boldsymbol{v},
\end{align}
\item \emph{approximation of the zero norm minimization (AROM)}, that achieves approximate $L_0$ penalization \citep[see][]{Weston2003} by solving the following objective function iteratively. It initializes a vector $\boldsymbol{z}=(1,\dots,1)$, and at each successive step, it resets 
$\boldsymbol{z}$ as $\boldsymbol{z} \times \boldsymbol{\beta}$,  
\begin{align} 
\min_{\boldsymbol{\beta},\beta_0} \sum_{i=1}^n q_i\max\{1 - \ys_i (\left\langle z_i*X_i, \boldsymbol{\beta}\right\rangle+\beta_0), 0\}/n +  \lambda \sum_{j=1}^p |\beta_j|^2
\end{align}
where $\boldsymbol{x} \ast \boldsymbol{w}= \left(x_1w_1,\dots,x_p w_p\right)$.

\item \emph{$L_0$ weighted support vector machines ($L_0$ WSVM)}, built on the works of \citet{Huang2010}, that achieves $L_0$ norm in weighted support vector machines through an iterative scheme, by solving the following objective function at the $t^{th}$ step,
\begin{align}
(\boldsymbol{\beta}^{(t)}, \beta_0^{(t)}) =\arg\min_{\boldsymbol{\beta},\beta_0} \sum_{i=1}^n q_i \xi_i/n +  \lambda \boldsymbol{\beta}^T \Lambda^{(t-1)}\boldsymbol{\beta},
\end{align}
subject to $\ys_i(\left\langle X_i \boldsymbol{\beta} \right\rangle +\beta_0) \geq 1-\xi_i,\:\xi_i\geq0,\:i=1,\dots,n$, where $\Lambda^{(t-1)} = diag(1/|\beta^{(t-1)}_1|^2,\dots,1/|\beta^{(t-1)}_p|^2)$.
\end{enumerate}

\tikzstyle{decision} = [diamond, draw,  
    text width=5em, text badly centered, node distance=3cm, inner sep=0pt]
\tikzstyle{block} = [rectangle, draw,  
    text width=8em, text centered, rounded corners, minimum height=2em]
\tikzstyle{line} = [draw, -latex']
\tikzstyle{cloud1} = [draw, ellipse, text badly centered, node distance=2cm,
    text width=3em, minimum height=1.5em]
\tikzstyle{cloud2} = [draw, ellipse, text badly centered, node distance=6.5cm,
    text width=10em, minimum height=2em]    

\begin{figure} 
\begin{center}
\resizebox{300pt}{!}{%
\begin{tikzpicture}[node distance = 2.5cm, auto]
    \node [block] (x1) {Calculate the empirical criterion $\mathcal{R}_{1}$ in the model without $\mathcal{X}_{(1)}$};
    \node [block, below of=x1,node distance=2cm] (xj1) {- - - - - -};
    \node [block, below of=xj1,node distance=0.8cm] (xj2) {- - - - - -};
    \node [block, below of=xj2,node distance=0.8cm] (xj3) {- - - - - -};
    \node [block,below of=xj3,node distance=2cm] (xdk1) {Calculate the empirical criterion $\mathcal{R}_{d-k}$ in the model without $\mathcal{X}_{(d-k)}$};
    \node [block,left of=xj2,node distance=4.5cm,text width=9em] (rem) {features left $\mathcal{Z}_{k} =\{\mathcal{X}_{(1)},\dots,\mathcal{X}_{(d-k)}\}$, features to remove $=\mathcal{S}_{k}$ };
    \node [block,left of=rem,node distance=4.5cm,text width=9em](init){Start with features left $\mathcal{Z}_{0}=\{\mathcal{X}_{1},\dots,\mathcal{X}_{d}\}$, features to remove $\mathcal{S}_{0}=\{0\}$.};
    \node [block,node distance=4.5cm,right of=xj2] (find) {Find feature $\mathcal{X}_{i_{k+1}}$ which produces lowest regularized empirical risk $\mathcal{R}_{i_{k+1}}$ };
    \node [decision,below of=find,node distance=4.5cm] (yesno) {Is $\mathcal{R}_{i_{k+1}} - \mathcal{R}_{i_{k}}$ too large? };
    \node [cloud1,node distance=2.5cm,below of=yesno] (stop) {Stop };
    \node [cloud2,left of=yesno,node distance=11.25cm,text width=12em] (update) {Update features left $\mathcal{Z}_{k+1} =\{\mathcal{X}_{(1)},\dots,\mathcal{X}_{(d-k-1)}\}$, update features to remove $\mathcal{S}_{k+1}=\mathcal{S}_{k}\cup \mathcal{X}_{i_{k+1}}$};
    \node [block,left of=stop,text width=9em, node distance=3.5cm] (output) {Features left $\mathcal{Z}_{k}$, and features to remove $\mathcal{S}_{k}$};
    \path [line] (init) -- node[pos=.55] (aux) {}(rem);
    \path [line] (rem) -- (x1);
    \path [line] (rem) -- (xj1);
    \path [line] (rem) -- (xj2);
    \path [line] (rem) -- (xj3);
    \path [line] (rem) -- (xdk1);
    \path [line] (x1) -- (find);
    \path [line] (xj1) -- (find);
    \path [line] (xj2) -- (find);
    \path [line] (xj3) -- (find);
    \path [line] (xdk1) -- (find);
    \path [line] (find) -- (yesno);
    \path [line] (yesno) --   node {yes}(stop);
    \path [line] (yesno) --   node {no} (update);
    \path [line] (update) -- (aux);
    \path [line] (stop) -- (output);
\end{tikzpicture}}
\end{center}
\caption{\textbf{Schematics of riskRFE in nonparametric estimation}}
\label{flow:rfe}
\end{figure}
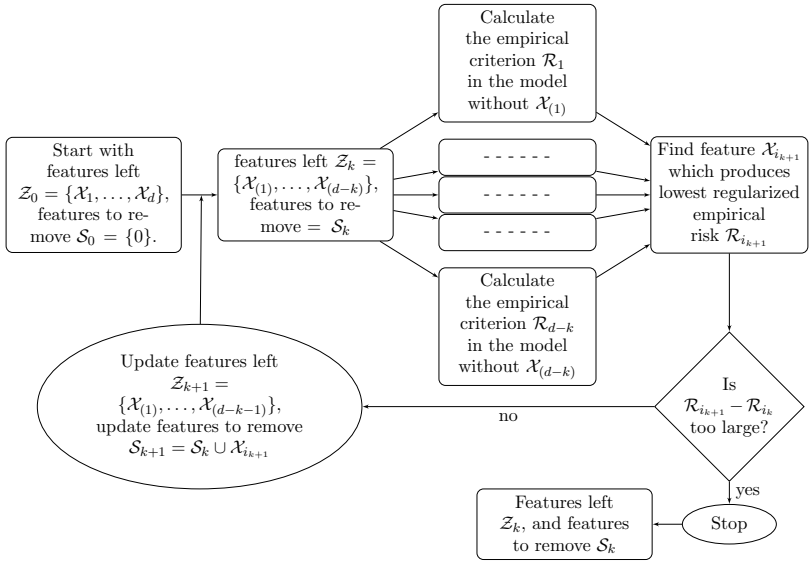

\subsection{Identifying optimal biomarker combinations in the nonlinear space.} 
In this section, we consider the derivation of best biomarker combinations in the nonlinear space, as most optimal marker combinations for treatment-selection rules may not be among the linear combinations of the input markers. 
Feature selection is a fairly straightforward procedure for linear SVM classifiers, but it is a much more challenging problem in nonlinear support vector machines \citep[see][]{Mangasarian2007}. For example, in the linear support vector machines, we can use the $L_1$ or $L_0$ norm instead of the Hilbert norm. However, when similar techniques are applied to nonlinear SVM classifiers, we have reduction in the number of support vectors but not in the number of input space features \citep{Fung2004}. So although the dimensionality of the transformed space is reduced, it does not provide any direct reduction of input space features. A few procedures have recently been developed however to deal with feature selection in the original input space under nonlinear feature maps \citep[see][]{Dasgupta2013, Allen2013}, but not a lot of them have been developed with the goal of $L_0$ penalization. We will investigate some of these newly developed methods here, but under the more generalized setting of weighted support vector machines. We briefly introduce each method below, along with their
modified objective function under the weighted SVM setup, while a more detailed description of each is reserved for the Supplementary Section (see Web Appendix \ref{sec:appB}). 
\begin{enumerate}
[leftmargin=0.2in,noitemsep]
\item \emph{risk recursive feature elimination (riskRFE)}, a newly developed, powerful wrapper technique based on recursive computation of the learning function \citep[see][]{Dasgupta2013}, that works both in the linear and the nonlinear space (see Figure \ref{flow:rfe} for a flow chart of the algorithm). 
\item \emph{kernel iterative feature extraction (KNIFE)}, that achieves feature selection in nonlinear SVMs, by optimizing a feature-regularized loss function (by weighting features within the kernel), iteratively \citep[see][]{Allen2013}. For $k_{\boldsymbol{w}}\left(\tx_1,\tx_2\right)=k\left(\boldsymbol{w}\ast \tx_1,\boldsymbol{w}\ast \tx_2\right)$, the weighted version of the algorithm solves, 
\begin{align}
\min_{\beta_0,\boldsymbol{\alpha}, \boldsymbol{w}} &\sum_{i=1}^n q_i\max\{1 - \ys_i (\sum_{j=1}^n\alpha_j k_{\boldsymbol{w}}(\tx_i,\tx_j)+\beta_0), 0\}/n +  \lambda_1 \alpha^T k_{\boldsymbol{w}} \alpha +\lambda_2\|\boldsymbol{w}\|_1 \nonumber\\
\text{subject to }& 0\leq w_j <1\text{ for all }j=1,\dots,p.
\end{align}
\item \emph{kernel penalized weighted support vector machines (KP-WSVM)}, built on the works of \citet{Maldonado2011} for feature selection in nonlinear SVMs, that relies on penalizing each feature's use in the dual formulation.
\begin{align}
 \min_{\beta_0,\boldsymbol{\alpha}, \boldsymbol{w}}&\sum_{i=1}^n q_i\max\{1 - \ys_i(\sum_{j=1}^n\alpha_j k_{\boldsymbol{w}}(\tx_i,\tx_j)+\beta_0), 0\}/n 
+  \lambda \sum_{j=1}^n (1 - e^{-\beta w_j}) \nonumber\\
\quad  \text{subject to }&w_j\geq 0 \quad \forall j=\{1,\dots,n\}.
\end{align}
The approximated $L_0$ penalty in the above objective function can also be replaced by a $L_1$ norm.
\end{enumerate}



\section{A simulation study}\label{sec:simu} In this section, we investigate the performance of the aforementioned feature selection methods in conjunction with the weighted support vector machines algorithms in minimizing the $L_0$ penalized sum of the weighted classification loss as in \eqref{eq:wt_sum2} in various simulation settings that we describe below.

\subsection{The simulation setup} We consider data from 1:1 randomized trial of size $n = 500$ for studying the performance of our proposed strategy. The linear feature selection methods (riskRFE (L), $L_1$ WSVM, SCAD/eSCAD, FSV, AROM, $L_0$ WSVM) are used in conjunction with the linear kernel based weighted SVM, while the nonlinear marker selection methods (riskRFE (G), KNIFE, $L_0$ KP-WSVM, $L_1$ KP-WSVM) are used in conjunction with the Gaussian RBF kernel based weighted SVM. For each simulation setting, we restrict ourselves to when the treatment/disease burden ratio $\delta_1$ is equal to 0, but consider three different values for the marker cost $\delta_2$, which is either $= 0/0.0001/0.001$ units. The feature selection methods considered here can be broadly categorized into two groups - (a) Embedded (penalized) methods, which have a separate tuning parameter for penalizing the $L_1$ or $L_0$ cost, including SCAD, eSCAD, $L_1$ WSVM, $L_0$ WSVM, FSV, KNIFE and KP-WSVMs. For these methods, $\delta_2$, the ratio of the cost of measuring one biomarker to the burden per disease event, is utilized directly for choosing the optimum value of the respective tuning parameter(s). Although AROM also belongs to this class, it achieves $L_0$ penalization through an iterative fitting of the $L_2$ norm SVM. Thus in this case, $\delta_2$ is used for choosing the optimum amount of $L_2$ tuning; and (b) Wrappers that have no separate penalization for the $L_1$ or $L_0$ cost (both riskRFE(L) and riskRFE(G)); thus are not tuned on the value of $\delta_2$. The other global SVM parameters, such as the width of the Gaussian kernel in nonlinear SVMs, are tuned based on the prediction performance of the weighted support vector machines in the full model. In this simulation study, we only consider the double-robustness weights of Huang and Fong (2014) for obtaining $W_i$ for each individual.
As a comparative method, we use linear logistic regression model with LASSO and weighted SVM without feature selection to find the optimal treatment selection rule. In logistic regression, the $P(Y=1|\tX,T)$ is modeled as a function of the main effects of the treatment and the biomarkers, and the interaction effects between the treatment and the biomarkers, with the amount of $L_1$ penalization tuned globally through usual cross validation. This is also the risk model used for constructing the double-robustness weights.

\begin{table}
\caption{\textbf{Setting 1(i) (Linear): Setting with disease prevalence approximately 0.30 and 0.20 among the untreated and treated, respectively (Number of markers 27, number of significant markers 2)} - Performance scores: (a) Monte Carlo mean of proportion of correct markers chosen, (b) Monte Carlo mean of proportion of incorrect markers chosen, (c) Monte Carlo mean of $\theta$ with (i) $\delta_1=0$ and $\delta_2=0$, (ii) with $\delta_1=0$ and $\delta_2=0.0001$ (iii) with $\delta_1=0$ and $\delta_2=0.001$ for different feature selection methods .\label{tab:set1_linear}}
\centering
\begin{threeparttable}
{\tabcolsep=0pt
\begin{tabular}{l ccccccccc}
\hline
\multicolumn{10}{c}{\textbf{Setting 1 (Underlying model: Linear)}}\\
\hline
& \multicolumn{3}{c}{$\delta_1=0$, $\delta_2=0$}&\multicolumn{3}{c}{$\delta_1=0$, $\delta_2=0.0001$}&\multicolumn{3}{c}{$\delta_1=0$, $\delta_2=0.001$}\\
Linear Methods&$\mathbb{E}$(Prop. & $\mathbb{E}$(Prop.  & $\mathbb{E}(\theta)$ &$\mathbb{E}$(Prop. & $\mathbb{E}$(Prop.  &$\mathbb{E}(\theta)$ &$\mathbb{E}$(Prop. & $\mathbb{E}$(Prop.  &$\mathbb{E}(\theta)$\\
&correct)&incorrect)&&correct)&incorrect)&&correct)&incorrect)&\\
\hline\\
SCAD &1&0.17&0.1048&1&0.16&0.1069&1&0.01&0.1045\\
elastic SCAD &1&0.20&0.1076&1&0.18&0.1128&1&0.11&0.1145\\
$L_1$ WSVM &1&0.31&0.1107&1&0.29&0.1123&1&0.16&0.1109\\
$L_0$ WSVM &1&0.26&0.1114&1&0.16&0.1105&1&0.08&0.1127\\
FSV &1&0.32&0.1074&1&0.22&0.1096&1&0.11&0.1075\\
AROM &0.97&0.10&0.1059&0.97&0.09&0.1079&0.97&0.05&0.1116\\
&&&&&&&&&\\
riskRFE (L) &1&0.03&0.1016&1&0.03&0.1019&1&0.03&0.1044\\
&&&&&&&&&\\
LASSO &1&0.29&0.1065&1&0.29&0.1074&1&0.29&0.1158\\ 
&&&&&&&&&\\
W-SVM Linear no sel.\tnote{\dag} &1&1&0.1178&1&1&0.1205&1&1&0.1448\\
&&&&&&&&&\\
\hline
& \multicolumn{3}{c}{$\delta_1=0$, $\delta_2=0$}&\multicolumn{3}{c}{$\delta_1=0$, $\delta_2=0.0001$}&\multicolumn{3}{c}{$\delta_1=0$, $\delta_2=0.001$}\\
Nonlinear Methods&$\mathbb{E}$(Prop. & $\mathbb{E}$(Prop.  & $\mathbb{E}(\theta)$ &$\mathbb{E}$(Prop. & $\mathbb{E}$(Prop.  &$\mathbb{E}(\theta)$ &$\mathbb{E}$(Prop. & $\mathbb{E}$(Prop.  &$\mathbb{E}(\theta)$\\
&correct)&incorrect)&&correct)&incorrect)&&correct)&incorrect)&\\
\hline\\
KNIFE&1& 0.27&0.1075&1&0.26&0.1101&1&0.12&0.1120\\
$L_1$ KP-WSVM &1&0.24&0.1080&1&0.24&0.1094&1&0.09&0.1082\\
$L_0$ KP-WSVM &1&0.32&0.1113&1&0.27&0.1102&1&0.17&0.1127\\
&&&&&&&&&\\
riskRFE (G) &1&0.03&0.1021&1&0.03& 0.1024&1&0.03&0.1049\\
&&&&&&&&&\\
W-SVM Gauss no sel.\tnote{\dag} &1&1&0.1202&1&1&0.1229&1&1&0.1472\\
&&&&&&&&&\\
\hline
\end{tabular}}
\begin{tablenotes}
\item[\dag] no sel. - no selection
\end{tablenotes}
\end{threeparttable}%
\end{table}

In each Monte Carlo simulation, we follow a 5-fold cross-validation procedure to identify the optimal tuning parameters in the model for a given method. As mentioned before, we perform this tuning in a two-step procedure - the global SVM parameters are tuned first using the full model in an weighted SVM analysis. The Hilbert Space norm of the estimated function (which in linear kernel becomes the $L_2$ norm) is tuned on a grid of values lying between $(0.0001,100)$, while the width of the Gaussian kernel, $\gamma$, is tuned over a grid of values lying between $(0.001,10)$. Then for each of the embedded methods, we use cross validation to tune for the additional penalization parameter(s) based on the performance of the method in question, using the selected global SVM parameters from step one. Each of these parameters was tuned on a grid of a viable range of values, and the value obtaining the optimal GCV performance is chosen for that method.  
This two-step procedure is followed to gain computation time, and also because the $L_2$ norm allows shrinkage of the estimated effects, rather than controlling sparsity directly.

{\small\begin{table}
\caption{\textbf{Setting 1(ii) (Nonlinear): Setting with disease prevalence approximately 0.18 and 0.17 among the untreated and treated, respectively (Number of markers 27, number of significant markers 2)} - Performance scores: (a) Monte Carlo mean of proportion of correct markers chosen, (b) Monte Carlo mean of proportion of incorrect markers chosen, (c) Monte Carlo mean of $\theta$ with (i) $\delta_1=0$ and $\delta_2=0$, (ii) with $\delta_1=0$ and $\delta_2=0.0001$ (iii) with $\delta_1=0$ and $\delta_2=0.001$ for different feature selection methods.\label{tab:set3_nonlinear}}
\centering
\begin{threeparttable}
{\tabcolsep=0pt
\begin{tabular}{l ccccccccc}
\hline
\multicolumn{10}{c}{\textbf{Setting 1 (Underlying model: Nonlinear)}}\\
\hline
& \multicolumn{3}{c}{$\delta_1=0$, $\delta_2=0$}&\multicolumn{3}{c}{$\delta_1=0$, $\delta_2=0.0001$}&\multicolumn{3}{c}{$\delta_1=0$, $\delta_2=0.001$}\\
Linear Methods&$\mathbb{E}$(Prop. & $\mathbb{E}$(Prop.  & $\mathbb{E}(\theta)$ &$\mathbb{E}$(Prop. & $\mathbb{E}$(Prop.  &$\mathbb{E}(\theta)$ &$\mathbb{E}$(Prop. & $\mathbb{E}$(Prop.  &$\mathbb{E}(\theta)$\\
&correct)&incorrect)&&correct)&incorrect)&&correct)&incorrect)&\\
\hline\\
SCAD &0.91&0.42&0.1296&0.89&0.32&0.1292&0.85&0.15&0.1322\\
elastic SCAD &0.96& 0.36&0.1289&0.96&0.33&0.1291&0.95&0.23&0.1340\\
$L_1$ WSVM &0.94&0.43& 0.1281&0.94&0.41&0.1296&0.90&0.29&0.1358\\
$L_0$ WSVM &0.88&0.30& 0.1247&0.86&0.29&0.1269&0.82&0.16&0.1314\\
FSV &0.89&0.32&0.1274&0.90&0.26& 0.1266&0.88&0.19&0.1328\\
AROM &0.91&0.25&0.1269&0.92&0.23&0.1263&0.89&0.16&0.1317\\
&&&&&&&&&\\
riskRFE (L)&0.85&0.10&0.1223&0.85&0.10& 0.1228&0.85&0.10&0.1266\\
&&&&&&&&&\\
LASSO&0.79&0.19&0.1320&0.79&0.19&0.1326&0.79&0.19&0.1383\\ 
&&&&&&&&&\\
W-SVM Linear no sel.\tnote{\dag} &1&1&0.1349&1&1&0.1376&1&1&0.1619\\
&&&&&&&&&\\
\hline
& \multicolumn{3}{c}{$\delta_1=0$, $\delta_2=0$}&\multicolumn{3}{c}{$\delta_1=0$, $\delta_2=0.0001$}&\multicolumn{3}{c}{$\delta_1=0$, $\delta_2=0.001$}\\
Nonlinear Methods&$\mathbb{E}$(Prop. & $\mathbb{E}$(Prop.  & $\mathbb{E}(\theta)$ &$\mathbb{E}$(Prop. & $\mathbb{E}$(Prop.  &$\mathbb{E}(\theta)$ &$\mathbb{E}$(Prop. & $\mathbb{E}$(Prop.  &$\mathbb{E}(\theta)$\\
&correct)&incorrect)&&correct)&incorrect)&&correct)&incorrect)&\\
\hline\\
KNIFE&0.99&0.35&0.1156&0.97&0.30&0.1147&0.95&0.21&0.1196\\
$L_1$ KP-WSVM &0.96&0.33&0.1130&0.96&0.31&0.1126&0.93&0.20&0.1183\\
$L_0$ KP-WSVM&0.97&0.45&0.1191&0.97&0.42&0.1196&0.95&0.35&0.1245\\
&&&&&&&&&\\
riskRFE (G) &0.88&0.11&0.1178&0.88&0.11&0.1183&0.88&0.11&0.1222\\
&&&&&&&&&\\
W-SVM Gauss no sel.\tnote{\dag} &1&1&0.1356&1&1&0.1383&1&1&0.1626\\
&&&&&&&&&\\
\hline
\end{tabular}}
\begin{tablenotes}
\item[\dag] no sel. - no selection
\end{tablenotes}
\end{threeparttable}
\end{table}

The optimal treatment-selection rule $\hat{A}(\hat{\tX}_0)$, where $\hat{\tX}_0\subseteq \tX_p$ is the optimal set of markers chosen by a given method,
is then estimated from the training sample, after retuning the global SVM parameters in the model with the selected markers. To evaluate the performance of the estimated rule, a test set of $n = 5000$ is generated in each simulation run, based on which we estimate $\theta$ as $\hat{\theta}=\sum_{i=1}^n \left\{(1-T_i)\times Y_i \times (1-\hat{A}(\hat{\tX}_{0i}))/n_0+T_i\times Y_i \times \hat{A}(\hat{\tX}_{0i})/n_1 +\delta_1\hat{A}(\hat{\tX}_{0i})\right\}+\delta_2|\hat{\tX}_0| 
$. 
We evaluate the performance of each method over 100 Monte Carlo runs. In our results, we present the estimated version of the quantity $\theta$ that we evaluate from these Monte Carlo runs. 
We compare performance of different methods in the following three settings.

\begin{table}
\caption{\textbf{Setting 2(i) (Linear): Setting with disease prevalence approximately 0.22 among both the untreated and treated (Number of markers 53, number of significant markers 3)} - Performance scores: (a) Monte Carlo mean of proportion of correct markers chosen, (b) Monte Carlo mean of proportion of incorrect markers chosen, (c) Monte Carlo mean of $\theta$ with (i) $\delta_1=0$ and $\delta_2=0$ for different feature selection methods.\label{tab:set2_linear}}
\centering
\begin{threeparttable}
{\tabcolsep=0pt
\begin{tabular}{l ccccccccc}
\hline
\multicolumn{10}{c}{\textbf{Setting 2 (Underlying model: Linear)}}\\
\hline
& \multicolumn{3}{c}{$\delta_1=0$, $\delta_2=0$}&\multicolumn{3}{c}{$\delta_1=0$, $\delta_2=0.0001$}&\multicolumn{3}{c}{$\delta_1=0$, $\delta_2=0.001$}\\
Linear Methods&$\mathbb{E}$(Prop. & $\mathbb{E}$(Prop.  & $\mathbb{E}(\theta)$ &$\mathbb{E}$(Prop. & $\mathbb{E}$(Prop.  &$\mathbb{E}(\theta)$ &$\mathbb{E}$(Prop. & $\mathbb{E}$(Prop.  &$\mathbb{E}(\theta)$\\
&correct)&incorrect)&&correct)&incorrect)&&correct)&incorrect)&\\
\hline\\
SCAD &0.98&0.31&0.0557&0.97&0.30&0.0582&0.95&0.01&0.0545\\
elastic SCAD &1&0.30&0.0599&1&0.23&0.0588&1&0.13& 0.0642\\
$L_1$ WSVM &1&0.45&0.0605&1&0.33&0.0574&1&0.14&0.0634\\
$L_0$ WSVM &0.97&0.23&0.0563&0.96&0.17&0.0565&0.95&0.01&0.0542\\
FSV &0.95&0.36&0.0594&0.94&0.24&0.0627&0.95&0.12&0.0630\\
AROM &0.94&0.06&0.0562&0.94&0.04&0.0549&0.94&0.02& 0.0598\\
&&&&&&&&&\\
riskRFE (L)&0.97&0.02&0.0519&0.97&0.02&0.0523&0.97&0.02&0.0558\\
&&&&&&&&&\\
LASSO &1&0.17&0.0555&1&0.17&0.0567&1&0.17&0.0670\\ 
&&&&&&&&&\\
W-SVM Linear no sel. \tnote{\dag}&1&1&0.0686&1&1&0.0739&1&1&0.1216\\
&&&&&&&&&\\
\hline
& \multicolumn{3}{c}{$\delta_1=0$, $\delta_2=0$}&\multicolumn{3}{c}{$\delta_1=0$, $\delta_2=0.0001$}&\multicolumn{3}{c}{$\delta_1=0$, $\delta_2=0.001$}\\
Nonlinear Methods&$\mathbb{E}$(Prop. & $\mathbb{E}$(Prop.  & $\mathbb{E}(\theta)$ &$\mathbb{E}$(Prop. & $\mathbb{E}$(Prop.  &$\mathbb{E}(\theta)$ &$\mathbb{E}$(Prop. & $\mathbb{E}$(Prop.  &$\mathbb{E}(\theta)$\\
&correct)&incorrect)&&correct)&incorrect)&&correct)&incorrect)&\\
\hline\\
KNIFE&0.98&0.24&0.0557&0.96&0.16&0.0567&0.96&0.03&0.0561\\
$L_1$ KP-WSVM &0.98&0.19&0.0550&0.98&0.08&0.0539&0.97&0.03&0.0560\\
$L_0$ KP-WSVM&0.99&0.20&0.0606&0.99&0.17&0.0572&0.99&0.09&0.0630\\
&&&&&&&&&\\
riskRFE (G) &0.98&0.02&0.0505&0.98&0.02&0.0509&0.98&0.02&0.0545	\\
&&&&&&&&&\\
W-SVM Gauss no sel. \tnote{\dag}&1&1&0.0681&1&1&0.0734&1&1&0.1211\\
&&&&&&&&&\\
\hline
\end{tabular}}
\begin{tablenotes}
\item[\dag] no sel. - no selection
\end{tablenotes}
\end{threeparttable}
\end{table}


In the first setting, we have in total a list of 27 markers, of which only 2 are significant in explaining treatment-marker interactions ($|\tX_p|=27,\:|\tX_0|=2$). The significant markers, $(X_1, X_2)$ are generated from the multivariate normal distribution $\sim N\left(\left(\begin{array}{c}
	0\\
	0
\end{array}\right)
, 
\left(\begin{array}{cc}
1 & 0.2\\
0.2 & 1	
\end{array}\right)
\right)$. Each of the rest is generated independently from a $N(0,1)$ distribution. We consider two subcases under this setting: (i) a `linear' underlying model, given by $logit P(Y = 1|X_1, X_2, T) = -1.5 - 1.5X_1 -1.25X_2 + 2X_1T + 1.5X_2T$, with disease prevalence approximately 0.3 and 0.2 among the untreated and treated, respectively; and (ii) a polynomial `nonlinear' underlying model
$logit P(Y = 1|X_1, X_2, T) = -1.5 + 0.2X_1 -0.2X_2 -3T - X_1T - X_2T+X_1^2T+X_2^2T$, with disease prevalence approximately 0.18 and 0.17 among the untreated and treated, respectively.

In the second setting, we have in total a list of 53 markers, of which only 3 are significant in explaining treatment-marker interactions ($|\tX_p|=53,\:|\tX_0|=3$). The significant markers, $(X_1, X_2, X_3)$ are generated from the multivariate normal distribution $ \sim N\left(
\left(\begin{array}{c}
	0\\
	0\\
	0
\end{array}\right)
, 
\left(\begin{array}{ccc}
1 & 0.2 &0.2\\
0.2 & 1	& 0.2\\
0.2 & 0.2 & 1
\end{array}\right)
\right)$. Each of the rest is generated independently from a $N(0,1)$distribution with mean 0 and standard deviation 1. Again we consider two subcases: (i) a `linear' underlying model, given by 
$logit P(Y = 1|X_1, X_2, X_3, T) = -2 +X_1+0.75X_2+X_3 - 2X_1T - 1.5X_2T - 2X_3T$, with disease prevalence approximately 0.22 among both the untreated and treated groups; and (ii) a `nonlinear' underlying model $logit P(Y = 1|X_1, X_2, X_3, T) = -0.8-3T - X_1T - X_2T+X_1^2T+X_2^2T-\sqrt{|X_3|}T$, with disease prevalence approximately 0.30 and 0.18 among the untreated and treated, respectively.

In the third setting, we consider a complex nonlinear setup, with a total of 27 markers, of which only 2 are significant in explaining treatment-marker interactions. All of the markers are generated from the uniform distribution, that is, $X_i \sim U(-2,2), \:i=1,\dots,27$. The model for treatment marker interaction considers a situation where the treatment has benevolent effect for an individual only if his/her marker values $X_1$ and $X_2$ lie in a specific region of the covariate space, without which the treatment might yield a harmful result. This is achieved by dividing the covariate space spanned by $X_1$ and $X_2$, the square $\{-2 \leq X_1\leq 2, -2 \leq X_2\leq 2\}$, into two regions: (i) a `harmful' zone given by the smaller square $\{-1 \leq X_1\leq 1, -1 \leq X_2\leq 1\}$, denoted as $H$; and (ii) a `beneficial' zone given by the region between the two concentric squares, denoted as $B$ (Figure \ref{fig:simu_set3} plots $X_1$ and $X_2$ marker values for 10000 such hypothetical individuals). Thus, the probability that the treatment is benevolent for the population is 0.75. The probability of disease for an individual in region $B$ is 0 if the individual receives treatment, but 0.6 otherwise, and they are reversed for an individual from $H$. The global probability for disease in the population is thus fixed at 0.3. The disease prevalences are approximately around 0.53 and 0.17 among the untreated and treated, respectively.

\begin{figure}
\centering
\includegraphics[width=0.6\textwidth]{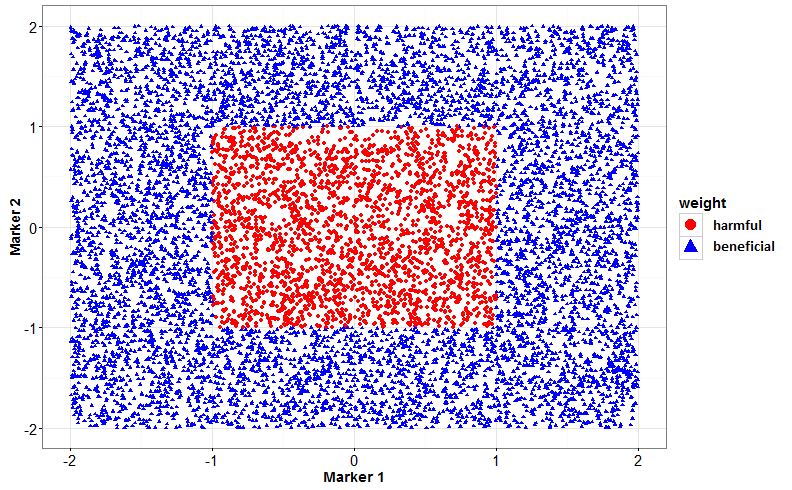}
\caption{\textbf{Simulation setting 3 -} Distribution of $X_1$ and $X_2$ marker values for 10000 individuals, stratified by their membership to the potential harm (red) or the potential benefit (blue) zones.}
\label{fig:simu_set3}
\end{figure}

\subsection{Results}\label{sec:res} The results for the simulation exercise are summarized in Tables \ref{tab:set1_linear}-\ref{tab:set5_nonlinear}. We give an overview of these results below, while a more detailed discussion of each of these settings is provided in the Supplementary Materials (see Web Appendix \ref{sec:appC}). 
\begin{enumerate}[leftmargin=0.2in,noitemsep]
\item While most biomarker-based treatment-selection rules result in reduction of the total cost compared to the optimal strategy between treating all or treating none, methods that allow for feature selection lead to substantial improvement compared to the weighted SVM method without feature selection. 
\item Embedded feature selection methods that involve tuning by marker measure cost $\delta_2$ in general show an decreasing trend in sensitivity and an increasing trend in specificity with increasing values of $\delta_2$, which is not the case for wrapper methods (such as riskRFE) or LASSO, that do not rely on $\delta_2$ for tuning. As a result, for the latter type of methods, the total cost $E(\theta)$ necessarily increases with increasing $\delta_2$, the cost of measuring a marker; in contrast, for methods that involve $\delta_2$ in tuning, the total cost $E(\theta)$ can often decrease with increasing $\delta_2$, especially when the improvement in the specificity score for a given method is more substantial than the decline in its sensitivity score.
\item In presence of a linear trend, best performing methods are typically among feature selection methods with linear kernel, yet feature selection methods with nonlinear kernel can have comparable performance; in presence of nonlinear trends, feature selection methods with nonlinear kernel can have substantial improvement over linear methods.
\item Relative performance of various methods vary with settings. In general, riskRFE performs really well when the cost of marker measurement is low. Apart from it, $L_1$ WSVM, SCAD and AROM are the best performing linear methods, especially when the cost of marker measurement is high, while $L_1$ KP-WSVM stands out among the nonlinear feature selection methods, with robust performance across settings.   
\end{enumerate}
We also evaluate these algorithms in a few additional settings. We compare their performance against that of the Decision List method of \citet{Zhang2015}. We also consider a setting where the true optimal decision rule is not sparse but the effect of some biomarkers on the optimal treatment decision are so small that given the cost consideration, a sparser decision rule is more optimal in terms of the criterion $\theta$. For space constraint, these additional settings are discussed in detail in the Supplementary Materials (Web Appendix \ref{sec:appD}), and the results are provided in Supplementary Tables \ref{tab:set4_decision}, \ref{tab:set4_linear} and \ref{tab:set5_linear}.}

\begin{table}
\caption{\textbf{Setting 2(ii) (Nonlinear): Setting with disease prevalence approximately 0.30 and 0.18 among the untreated and treated, respectively (Number of markers 53, number of significant markers 3)} - Performance scores: (a) Monte Carlo mean of proportion of correct markers chosen, (b) Monte Carlo mean of proportion of incorrect markers chosen, (c) Monte Carlo mean of $\theta$ with (i) $\delta_1=0$ and $\delta_2=0$, (ii) with $\delta_1=0$ and $\delta_2=0.0001$ (iii) with $\delta_1=0$ and $\delta_2=0.001$ for different feature selection methods.\label{tab:set4_nonlinear}}
\centering
\begin{threeparttable}
{\tabcolsep=0pt
\begin{tabular}{l ccccccccc}
\hline
\multicolumn{10}{c}{\textbf{Setting 2 (Underlying model: Nonlinear)}}\\
\hline
& \multicolumn{3}{c}{$\delta_1=0$, $\delta_2=0$}&\multicolumn{3}{c}{$\delta_1=0$, $\delta_2=0.0001$}&\multicolumn{3}{c}{$\delta_1=0$, $\delta_2=0.001$}\\
Linear Methods&$\mathbb{E}$(Prop. & $\mathbb{E}$(Prop.  & $\mathbb{E}(\theta)$ &$\mathbb{E}$(Prop. & $\mathbb{E}$(Prop.  &$\mathbb{E}(\theta)$ &$\mathbb{E}$(Prop. & $\mathbb{E}$(Prop.  &$\mathbb{E}(\theta)$\\
&correct)&incorrect)&&correct)&incorrect)&&correct)&incorrect)&\\
\hline\\
SCAD &0.84&0.65&0.1578&0.74&0.39&0.1588&0.68&0.24&0.1693\\
elastic SCAD &0.88&0.63&0.1607&0.81&0.47&0.1641&0.77&0.39&0.1821\\
$L_1$ WSVM &0.76&0.47&0.1597&0.71&0.37&0.1610&0.58&0.18&0.1696\\
$L_0$ WSVM &0.74&0.33&0.1576&0.69&0.27&0.1621&0.63&0.15&0.1685\\
FSV &0.83&0.64&0.1597&0.70&0.40&0.1630&0.69&0.33&0.1762\\
AROM &0.74&0.33&0.1564&0.68&0.24&0.1584&0.65&0.17&0.1653\\
&&&&&&&&&\\
riskRFE (L)&0.64&0.22&0.1564&0.64&0.22&0.1577&0.64&0.22&0.1693\\
&&&&&&&&&\\
LASSO&0.59&0.15&0.1636&0.59&0.15&0.1645&0.59&0.15&0.1729\\ 
&&&&&&&&&\\
W-SVM Linear no sel.\tnote{\dag} &1&1&0.1674&1&1&0.1727&1&1&0.2204\\
&&&&&&&&&\\
\hline
& \multicolumn{3}{c}{$\delta_1=0$, $\delta_2=0$}&\multicolumn{3}{c}{$\delta_1=0$, $\delta_2=0.0001$}&\multicolumn{3}{c}{$\delta_1=0$, $\delta_2=0.001$}\\
Nonlinear Methods&$\mathbb{E}$(Prop. & $\mathbb{E}$(Prop.  & $\mathbb{E}(\theta)$ &$\mathbb{E}$(Prop. & $\mathbb{E}$(Prop.  &$\mathbb{E}(\theta)$ &$\mathbb{E}$(Prop. & $\mathbb{E}$(Prop.  &$\mathbb{E}(\theta)$\\
&correct)&incorrect)&&correct)&incorrect)&&correct)&incorrect)&\\
\hline\\
KNIFE&0.79&0.31&0.1435&0.74&0.21&0.1388&0.71& 0.07&0.1304\\
$L_1$ KP-WSVM &0.81&0.53&0.1528&0.62&0.27&0.1513&0.50&0.10&0.1455\\
$L_0$ KP-WSVM&0.82&0.53&0.1535&0.81&0.47&0.1567&0.77&0.38&0.1642\\
&&&&&&&&&\\
riskRFE (G) &0.65&0.09&0.1369&0.65&0.09&0.1375&0.65&0.09&0.1432\\
&&&&&&&&&\\
W-SVM Gauss no sel.\tnote{\dag} &1&1& 0.1645&1&1&0.1698&1&1&0.2175\\
&&&&&&&&&\\
\hline
\end{tabular}}
\begin{tablenotes}
\item[\dag] no sel. - no selection
\end{tablenotes}
\end{threeparttable}
\end{table}

\begin{table}
\caption{\textbf{Setting 3 (Nonlinear): Setting with disease prevalence approximately 0.53 and 0.17 among the untreated and treated, respectively (Number of markers 27, number of significant markers 2)} - Performance scores: (a) Monte Carlo mean of proportion of correct markers chosen, (b) Monte Carlo mean of proportion of incorrect markers chosen, (c) Monte Carlo mean of $\theta$ with (i) $\delta_1=0$ and $\delta_2=0$ for different feature selection methods.\label{tab:set5_nonlinear}}
\centering
\begin{threeparttable}
{\tabcolsep=0pt
\begin{tabular}{l ccccccccc}
\hline
\multicolumn{10}{c}{\textbf{Setting 3 (Underlying model: Nonlinear)}}\\
\hline
& \multicolumn{3}{c}{$\delta_1=0$, $\delta_2=0$}&\multicolumn{3}{c}{$\delta_1=0$, $\delta_2=0.0001$}&\multicolumn{3}{c}{$\delta_1=0$, $\delta_2=0.001$}\\
Linear Methods&$\mathbb{E}$(Prop. & $\mathbb{E}$(Prop.  & $\mathbb{E}(\theta)$ &$\mathbb{E}$(Prop. & $\mathbb{E}$(Prop.  &$\mathbb{E}(\theta)$ &$\mathbb{E}$(Prop. & $\mathbb{E}$(Prop.  &$\mathbb{E}(\theta)$\\
&correct)&incorrect)&&correct)&incorrect)&&correct)&incorrect)&\\
\hline\\
SCAD &NA&NA&NA&NA&NA&NA&NA&NA&NA\\
elastic SCAD &NA&NA&NA&NA&NA&NA&NA&NA&NA\\
$L_1$ WSVM &0.19&0.18&0.1744&0.17&0.20&0.1749&0.16&0.19&0.1795\\
$L_0$ WSVM &1&0.99&0.1746&1&1&0.1773&1&1&0.2015\\
FSV &0.91&0.95&0.1745&0.50&0.52&0.1746&0.50&0.52&0.1873\\
AROM &NA&NA&NA&NA&NA&NA&NA&NA&NA\\
&&&&&&&&&\\
riskRFE (L)&0.22&0.29&0.1749&0.22&0.29&0.1757&0.22&0.29&0.1826\\
&&&&&&&&&\\
LASSO&0.05&0.14&0.1776&0.05&0.14&0.1780&0.05&0.14&0.1812\\ 
&&&&&&&&&\\
W-SVM Linear no sel.\tnote{\dag} &1&1&0.1746&1&1&0.1773&1&1&0.2015\\
&&&&&&&&&\\
\hline
& \multicolumn{3}{c}{$\delta_1=0$, $\delta_2=0$}&\multicolumn{3}{c}{$\delta_1=0$, $\delta_2=0.0001$}&\multicolumn{3}{c}{$\delta_1=0$, $\delta_2=0.001$}\\
Nonlinear Methods&$\mathbb{E}$(Prop. & $\mathbb{E}$(Prop.  & $\mathbb{E}(\theta)$ &$\mathbb{E}$(Prop. & $\mathbb{E}$(Prop.  &$\mathbb{E}(\theta)$ &$\mathbb{E}$(Prop. & $\mathbb{E}$(Prop.  &$\mathbb{E}(\theta)$\\
&correct)&incorrect)&&correct)&incorrect)&&correct)&incorrect)&\\
\hline\\
KNIFE&0.42&0.69&0.1765&0.39&0.68&0.1799&0.40&0.71&0.1958\\
$L_1$ KP-WSVM &0.88&0.32&0.1100&0.88&0.28&0.1060&0.88&0.26&0.1127\\
$L_0$ KP-WSVM&1&0.88&0.1621&1&0.89&0.1667&1&0.88&0.1860\\
&&&&&&&&&\\
riskRFE (G) &0.37&0.08&0.1349&0.37&0.08&0.1352&0.37&0.08&0.1376\\
&&&&&&&&&\\
W-SVM Gauss no sel.\tnote{\dag} &1&1&0.1761&1&1&0.1788&1&1&0.2031\\
&&&&&&&&&\\
\hline
\end{tabular}}
\begin{tablenotes}
\item[\dag] no sel. - no selection
\end{tablenotes}
\end{threeparttable}
\end{table}

\section{Real data analysis}\label{sec:real_data}

We now use an example from the RV144 Thailand HIV vaccine trial to examine the performance of the methods discussed above for selecting and combining markers. RV144 is one of the first vaccine trials that showed a significant positive effect of vaccine in preventing HIV infection. It included 16,402 participants, aged between 18 and 30 years, randomized 1:1 between vaccine and placebo \citep{Rerks-Ngarm2009}. A followup hostgenetic study was conducted to measure the effect of genotypes of Fc receptor genes on vaccine efficacy, and to that effect 190 single nucleotide polymorphisms (SNPs) (including five Fc-$\gamma$ and one Fc-$\alpha$ receptors) were genotyped on 125 cases (74 placebo recipients and 51 vaccine recipients), and 225 controls (20 placebo recipients and 205 vaccine recipients), of which 28 SNPs were selected based on Hardy-Weinberg equilibrium by \citet{Li2014}, each categorized into a binary variable, to study association of each with vaccine efficacy. Here we consider all 28 SNPs as the expensive candidate biomarkers to select from. In addition, age, gender, and baseline behaviorial risk for vaccine are combined with SNPs for treatment recommendation as in \citet{Huang2015}; no penalty is put on those baseline variables as they are readily available from the trial. Lifetime HIV treatment cost is estimated to be around \$370K according to CDC (\url{https://www.cdc.gov/hiv/programresources/guidance/costeffectiveness/index.html}). Vaccines typically cost a few hundred dollars and have minimal side effects, so we set $\delta_1=0.001$. The average cost of a SNP evaluation varies between \$ 2-4, and hence we consider a range of values for $\delta_2 \in \{10^{-6}, 5\times 10^{-6}, 10^{-5}\}$ based on the burden of measuring a SNP with respect to the treatment cost (as $2/370000 \approx 5\times 10^{-6}$). We compare the performance of the linear and nonlinear feature selection methods to select the optimal subset of markers for the purposes of vaccine recommendation for an individual. We also use LASSO as a method of selecting markers and estimating the treatment recommendation rule. A fivefold CV is performed to select $\lambda$, the parameter which controls for overfitting in the estimated WSVM decision function, and we do the same to select the width of the Gaussian kernel, whenever it was used. We also employed a fivefold CV to choose the other tuning parameters, specific to each of the feature selection methods employed. To compute the expected disease rate for marker selection for each of these methods, we perform a cross validation procedure by splitting the data into five random folds, and using four folds for training the WSVM (or LASSO) procedures with the selected markers, and using the remaining one for testing. The procedure is repeated 100 times and the average disease rate is computed.

\begin{table}
\caption{\textbf{Cross-validated treatment-selection performance of various approaches for making treatment recommendation in the RV144 trial:} Disease cost $+$ vaccine cost $+$ SNP cost - expected disease rate per 1000 individuals with added cost of vaccine ($\delta_1=0.001$) and 3 different costs for marker evaluation ($\delta_2\in \{10^{-6}, 5\times 10^{-6}, 10^{-5}\}$ for each of the 28 SNPs evaluated) for treatment selection with various methods of marker selection.\label{tab:realdat}}
\centering
\begin{threeparttable}
\begin{tabular}{l@{\hskip 0.05in}c@{\hskip 0.05in}c@{\hskip 0.05in}c}
&&&\\
\hline
 & Disease cost & Disease cost & Disease cost\\
Methods &  $+$ vaccine cost &  $+$ vaccine cost &  $+$ vaccine cost \\
&$+$SNP cost&$+$SNP cost&$+$SNP cost\\
&$\delta_2=10^{-6}$&$\delta_2=5\times10^{-6}$&$\delta_2=10^{-5}$\\
\hline
&&&\\
Treat all &7.44&7.44&7.44\\
Treat none &8.54&8.54&8.54\\
&&&\\
\hline
&&&\\
SCAD&7.21&7.28&7.37\\
elastic SCAD &7.34&7.44&7.56\\
$L_1$ WSVM &7.24&7.32&7.42\\
$L_0$ WSVM &7.31&7.37&7.44\\
FSV&7.29&7.35&7.43\\
AROM &6.97&7.03&7.11\\
riskRFE (L) &7.15&7.17&7.19\\
LASSO &7.53&7.56&7.60\\
W-SVM Linear no sel.\tnote{\dag} &7.42&7.55&7.71\\
&&&\\
\hline
&&&\\
KNIFE&7.33&7.38&7.44\\
$L_1$ KP-WSVM&7.39&7.46&7.55\\
$L_0$ KP-WSVM &7.32&7.42&7.55\\
riskRFE (G) &7.35&7.40&7.46\\
W-SVM Gauss no sel.\tnote{\dag}&7.44&7.57&7.73\\
&&&\\
\hline
\end{tabular}
\begin{tablenotes}
\item[\dag] no sel. - no selection
\end{tablenotes}
\end{threeparttable}
\end{table}

Table \ref{tab:realdat} shows the estimated performance of different selection strategies along with the strategy of treating none and treating all. Those two lead to an estimated HIV infection rate of 8.54 and 6.44 per 1000 persons, respectively, consistent with the positive vaccine efficacy we observed in the RV144 trial. From Table \ref{tab:realdat}, it is clear that linear weighted support vector machines (with or without marker selection) does a better job of treatment recommendation compared to the nonlinear weighted support vector machines (with or without marker selection) with the Gaussian kernel in this particular example. It can be seen that both the linear and nonlinear WSVM without any selection perform at par with the strategy of treating all, even when we consider a biomarker cost, when it is relatively low, but gets worse with higher values of $\delta_2$. Most of the linear selection methods yield a total cost at least as good as the strategy of treating all, even when $\delta_2$ is high. Among the linear methods, only elastic SCAD achieves a total cost worse than the strategy of treating all when $\delta_2$ is high. On the other hand, AROM is outperforming every other strategy, followed closely by riskRFE (L).
Overall we can conclude that marker selection is an important formulation for treatment recommendation using the weighted support vector machines framework.

\section{Concluding remarks}\label{sec:disc} In this article, we developed a new framework to incorporate marker measurement cost into treatment regime identification, with a pre-determined cost for inclusion of each biomarker in the model. We extended several different marker selection methods to apply to the weighted support vector machines setting, encompassing both the linear and the nonlinear space, in order to derive the optimal treatment-selection rules that minimize the total cost due to disease, treatment, and marker measurements. We investigated their performance in a number of different setups through a detailed simulation study, and also in a real data scenario, the RV144 HIV vaccine trial. We showed that in presence of a large number of candidate biomarkers, marker selection is essential for deriving cost-effective treatment-selection rules that effectively reduces disease and treatment burdens to the population while avoiding the burdens of collecting information on irrelevant biomarkers. We showed that marker selection also reduced the chance of obtaining overfitted treatment recommendation rules, which can result in lower test disease rate predictions than when we use the full model for treatment recommendation. It is worthwhile to note here that the indirect approaches can sometimes be inefficient compared with direct approaches, especially when the underlying disease risk model is correctly specified in the direct approaches. Moreover, inference for the estimated optimal regime can become challenging under indirect approaches. However, these methods are still appealing and provide a complementary alternative to direct approaches because of their robustness to model misspecifications. In fact, comparison of our algorithms with logistic regression with LASSO showed that in some settings, even when the risk model assumptions hold (as was the case in most of our simulation settings), feature selection using the indirect approaches can lead to treatment-selection rules with performance comparable to or even better than that based on direct approaches. Among the various marker selection technique employed in the simulations, riskRFE stood out as the best performing method when cost of measuring marker measurement cost was low, while AROM, SCAD and $L_0$ WSVM were the best performing linear feature selection methods for higher costs of marker selection, depending on the setting. On the other hand, $L_1$ KP-WSVM was clearly the best performing nonlinear feature selection method in most of the settings. In general, we showed that using nonlinear feature selection methods can perform substantially better than linear feature selection methods in presence of nonlinear patterns, at a minimal cost of efficiency in presence of linear patterns. One last thing to note here is that the methodology put forward in this article typically invoke the randomized setting. However, if the assumptions of (i) SUTVA and (ii) ignorable treatment assignment (see Page 5 in Section \ref{sec:methods} for more details) hold in an observational setting too, our proposed methods will continue to be applicable there, although assumption (ii) is typically not verifiable without randomization.

%

\section{Supplementary Materials}
\label{sec:supp}
Additional materials referenced in Sections~\ref{sec:methods} and \ref{sec:simu} are available with this paper at the Journal of Royal Statistical Society Series C web page hosted at the Wiley Online Library.

\section{Acknowledgments} The work was supported by NIH grant R01 GM106177-01. The authors thank the participants, investigators, and sponsors of the RV144 trials. 
The authors thank Dr. Dan Geraghty from the Fred Hutchinson Cancer Center for generating the genetics data, and Drs. Sue Li and Peter Gilbert for pre-processing and preliminary analysis of the SNP data. The views expressed are those of the authors and should not be construed to represent the positions of the U.S. Army or the Department of Defense.

\end{document}